# Hybrid deep learning architecture for general disruption prediction across tokamaks

J.X. Zhu[1], C. Rea[1], K. Montes[1], R.S. Granetz[1], R. Sweeney[1] and R.A. Tinguely[1]

[1] Plasma Science and Fusion Center, Massachusetts Institute of Technology, Cambridge, MA USA

E-mail: jxzhu@mit.edu

**Abstract**

In this paper, we present a new deep learning disruption prediction algorithm based on important findings from explorative data analysis which effectively allows knowledge transfer from existing devices to new ones, thereby predicting disruptions using very limited disruptive data from the new devices. The explorative data analysis conducted via unsupervised clustering techniques confirms that time-sequence data are much better separators of disruptive and non-disruptive behavior than the instantaneous plasma state data with further advantageous implications for a sequence-based predictor. Based on such important findings, we have designed a new algorithm for multi-machine disruption prediction that achieves high predictive accuracy on the C-Mod (AUC=0.801), DIII-D (AUC=0.947) and EAST (AUC=0.973) tokamaks with limited hyperparameter tuning. Through numerical experiments, we show that boosted accuracy (AUC=0.959) is achieved on EAST predictions by including in the training only 20 disruptive discharges, thousands of non-disruptive discharges from EAST, and *combining* this with more than a thousand discharges from DIII-D and C-Mod. The improvement of predictive ability obtained by combining disruptive data from other devices is found to be true for all permutations of the three devices. Furthermore, by comparing the predictive performance of each individual numerical experiment, we find that non-disruptive data are machine-specific while disruptive data from multiple devices contain device-independent knowledge that can be used to inform predictions for disruptions occurring on a new device.

## 1. Introduction

Utilizing nuclear fusion energy via magnetic-confinement tokamaks is one of a few encouraging paths toward future sustainable energy. Along the way, scientists need to learn to avoid plasma disruptions: these sudden and unexpected plasma terminations still represent one of the key challenges for tokamak devices, as their deleterious consequences can damage the whole fusion device and prevent the realization of a burning plasma reactor. Forecasting plasma instabilities and disruptions using first-principle models has demonstrated to be extremely difficult due to the complexity of the problem and the high non-linearity of the system [1]. To date, disruption prediction has been studied through two main approaches: data-driven versus physics-driven (or model-based). On one hand, recent statistical and machine learning (ML) approaches based on experimental data have shown attractive results for disruption prediction, even in real-time environments [2-10]. Different tokamak devices have different operational spaces, spatiotemporal scales for physics events and plasma diagnostics. Therefore, most of these data-driven approaches were developed and optimized specifically for one device and did not show promising cross-device prediction ability [2-6, 8, 10]. Specifically, cross-machine studies such as [7] focused on predictors that were trained on datasets purely or mostly from one device: these predictors achieved great performances on the training device but lacked the generalization capabilities derived from an understanding of the underlying physics, and therefore tended to fail on new, unseen device data. In addition, the complexity of these data-driven models limits their physics interpretability. Recently published work by Kates-Harbeck et al. [9] for the first time demonstrates the potential of predictors based on Deep Learning (DL) in learning a general representation of experimental data that can be used in cross-machine applications. On the other hand, model-based disruption prediction studies [11-15] seek to identify event chains that can lead to disruptions through an early event detection, which could help operators avoid disruptions. As an example, the Disruption Event Characterization and Forecasting (DECAF) [12] suite can detect a variety of events by taking advantage of first-principle, physics-based modules for tearing stability, resistive wall modes, etc. These models are tailored for stability limit detection on different devices and are sometimes accelerated through the adoption of surrogate ML models [16]; this latter approach tries to take advantage of both paradigms by incorporating physics and ML aspects in the same model to achieve better prediction as well as improved interpretability.

In this paper, we focus on the data-driven approach exploiting existing experimental databases for disruption prediction available on the Alcator C-Mod, DIII-D, and EAST tokamaks. We first introduce an application of a dimensionality reduction algorithm to the high-dimensional plasma data; then we discuss a new DL framework to disruption prediction based on important findings from the explorative data analysis which allows effective knowledge transfer from existing devices to new ones using very limited disruptive data from the new devices, while retaining high accuracy on the individual datasets. To this aim, we selected a set of disruption-relevant physical signals, available on all of the analyzed tokamak devices, and developed a powerful general algorithm using large databases from these three very different tokamaks. In addition, we combined data from the three different devices to add some randomization to the training domain which can alleviate over-learning of one specific device's behavior, and designed a set of cross-machine experiments to find general guidelines for disruption prediction on new devices using very limited disruptive data from themselves. Moreover, unlike previous studies [7, 9-10], the three machines have very different features: EAST is a medium-sized (R=1.85m, a=0.45m) superconducting tokamak with a hybrid first wall: its lower divertor is Carbon, the middle wall Molybdenum (Mo), while the upper divertor is made of Tungsten [17]. DIII-D is a medium-sized (R=1.67m, a=0.67m) tokamak with a carbon wall and relatively large error field: most of disruptive shots in our DIII-D database contain a locked mode as the last precursor in their event chain toward disruption [18-19]. C-Mod is a smaller tokamak (R=0.68m, a=0.22m) with high energy density (plasma pressure up to 2.05atm), high magnetic field ($B_T$ up to 8T) and high-Z metal (Mo) wall. The combination of these different characteristics covers a substantial fraction of ITER's features [20-21], though no existing device by itself can fully represent ITER at scale. A cross-machine study using data from these existing devices is nevertheless well-suited for investigating disruption prediction solutions for ITER.

## 2. Dataset description

Our disruption prediction studies are conducted on disruption warning datasets on three machines [8]: Alcator C-Mod (2012-2016 campaigns), DIII-D (2014-2018 campaigns) and EAST (2014-2018 campaigns). For all three databases, we include all types of disruptions except for intentional ones. The choice of which parameters to include in the databases is guided by our knowledge of the plasma physics mechanisms inherent to disruption characteristics of the different devices, as well as the accessibility and consistency of these parameters on all three machines. Many of the disruption-relevant parameters included in this study are also influenced by several papers [22-24]. The considered signals for the predictive models reported in this paper and their definitions can be found in table 1, while the composition of the three training datasets is shown in table 2. Given these databases, we formalize the disruption prediction problem in a *sequence-to-label* supervised machine learning framework, where we assign a label to each input plasma sequence, S (a 10-step consecutive sequence in time of 12 plasma signals) and train an algorithm to learn the functional representation, mapping the input sequences to one of two possible labels, 'disruptive' or 'non-disruptive'. To this aim, we explicitly defined different time thresholds for each machine to identify the unstable phase of the disruptive training discharges and assigned the disruptive label to plasma sequences that intersect the unstable phase of disruptive experimental runs, while the non-disruptive label is assigned to sequences extracted from the non-disruptive discharges. This classification scheme implicitly assumes that it is possible to detect a transition in time from a safe operational regime to a disruptive one and is another instance of incorporating physics knowledge into the Artificial Intelligence (AI) workflow [25-26]. The chosen time thresholds vary on different considered devices which depend on the transition points where some of plasma parameters exhibit identifiable changes in behavior when disruptions occur, considering a notable fraction of disruptions [8] and the suggestions from tokamak operators.

The training samples are ordered into sequences of ten time slices extracted from each shot of the training dataset. For each shot, we randomly select a subset of examples: this is a model's hyperparameter, tuned for each machine. The disruptive training sequences are randomly extracted from all sequences that intersect the unstable phase of each disruptive shot, while those sequences outside of the unstable region are not included in the training set. If disruptive patterns are learned properly, the algorithm will be able to identify similar trends also at times prior to the formally set time threshold, enabling the detection of early disruptive precursors. The non-disruptive sequences are randomly extracted from the flattop of non-disruptive training discharges. It is interesting to note that the database population consists of mostly non-disruptive data, thus resulting in a dataset imbalanced with respect to disruptive data.

**Table 1: Descriptions and symbols of all considered signals**

| Signal description | Symbol |
|---|---|
| $\dfrac{\text{Plasma current} - \text{programed plasma current}}{\text{Programed plasma current}}$ | `ip-error-fraction` |
| Perburbed field of nonrotating mode [a] <br> (n = 1 Fourier component), $B^{n=1}/B_{tor}$ | `locked-mode-proxy` |
| $\dfrac{\text{Electron density}}{\text{Greenwald density}}$ | `Greenwald-fraction` |
| Distance between the plasma and the lower divertor | `lower-gap` |
| Current centroid vertical position error [b] | `z-error-proxy` |
| Plasma elongation | `kappa` |
| Normalized plasma pressure (ratio of thermal to poloidal magnetic pressure) | `betap` |
| $\dfrac{\text{Radiated power}}{\text{Input power}}$ | `radiated-fraction` |
| Standard deviation of the magnetic field [c] measured from an array of Mirnov coils, normalized by $B_{tor}$ | `rotating-mode-proxy` |
| Loop voltage $V_{loop}$ | `v-loop` |
| Safety factor at the 95% flux surface | `q95` |
| Normalized internal inductance | `li` |

[a] For the C-Mod database, the `locked-mode-proxy` signal is obtained from a Mirnov coil array instead of the saddle coil.

[b] For the DIII-D database, we use current centroid vertical position instead of position error for the `z-error-proxy` signal.

[c] For the DIII-D database, we use n=1 component of magnetic field measured from a Mirnov coil array normalized by $B_{tor}$ for the `rotating-mode-proxy` signal.

**Table 2: The dataset composition of the three disruption warning databases**

|  | NO. training shots | NO. test shots | NO. validation shots | Sampling Rate (ms) | Time Threshold (ms) | NO. samples/training |
|---|---|---|---|---|---|---|
| C-Mod | 3343 (692) | 651 (142) | 463 (98) | 5 | 75 | 16 |
| DIII-D | 5286 (732) | 1085 (157) | 734 (107) | 10 | 400 | 25 |
| EAST | 8296 (2301) | 1674 (475) | 1137 (322) | 25 | 500 | 20 |

Values in parentheses give the exact number of disruptive shots.

## 3. Explorative data analysis through data visualization technique

As mentioned in the introduction, disruptions are highly complicated phenomena, characterized by diverse physics events with different spatiotemporal scales and non-linear dynamics [1]. In particular, we usually have to deal with high-dimensional data from multiple plasma signals which complicates both analysis and physics interpretation when studying disruptive events. In this section, we discuss the application of a nonlinear dimensionality reduction technique called **t**-Distributed **S**tochastic **N**eighbor **E**mbedding (t-SNE) [27] (see the appendix for further details of the t-SNE algorithm) to visualize high-dimentional plasma data in a 2-D plane to study the inherent data structure of the considered plasma signals. In principle, the t-SNE algorithm can be applied to any high-dimensional database. However, in this section, we only show the application to the C-Mod database as it is considered more difficult to predict through data-driven approach than EAST and DIII-D [6]. The analysis of DIII-D and EAST disruption databases can be found in the appendix.

Figure 1 shows the t-SNE algorithm applied to time slice data (left) and aggregated sequence data (right) for the C-Mod disruption warning database. In the left subplot, each blue point represents a randomly sampled time slice (a 12-element array composed by the 12 plasma signals from table 1) from the flattop of a non-disruptive shot, while each red point represents a time slice randomly sampled from the last 75 ms of a disruptive shot. On the right, each red point represents a 10-step (a 10 x 12 element matrix) sequence randomly sampled from the last 75 ms of a disruptive shot while each blue point represents a 10-step sequence randomly sampled from the flattop of a non-disruptive shot. We include all disruptions without discriminating by the cause. The coloring of each datapoint in the plots is done *a-posteriori*, i.e. not provided during the training process, thereby characterizing the t-SNE as an unsupervised clustering technique.

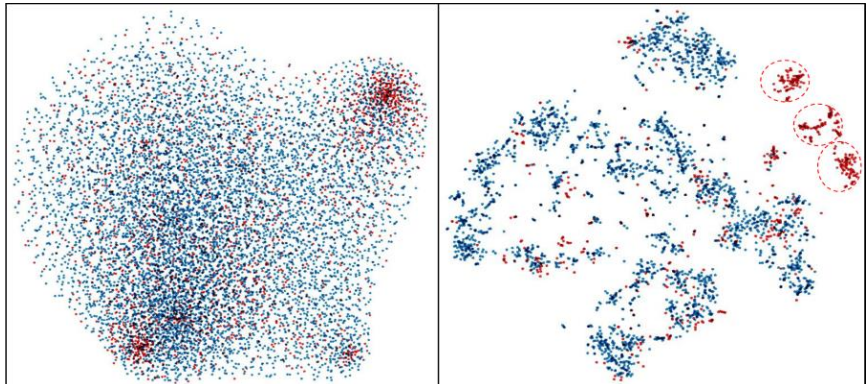

**Figure 1: t-SNE clustering for visualizing C-Mod data. On the left, t-SNE is performed on individual disruptive (red) and non-disruptive (blue) time slices while on the right t-SNE is performed on 10-step disruptive (red) and non-disruptive (blue) sequences. Three major clusters of disruptive data can be isolated. The colouring is done a-posteriori.**

Several important conclusions can be drawn: first, the clustering of individual time slices cannot isolate clear data clusters in the low-dimensional map. However, by performing t-SNE on 10-step plasma sequences, it is possible to isolate three major clusters - identified by dashed circles in figure 1 - which account for approximately 60% of the disruptive data. The improved separation of disruptive and non-disruptive data obtained when clustering sequences highlights the importance of temporal correlation and mutual information between consecutive time slices. This further suggests that sequence-based classifiers could have a clear advantage over the time slice based ones. Secondly, the t-SNE application to C-Mod sequences reveals that a substantial fraction (~40%) of disruptive sequences remains mixed with non-disruptive sequences. Further analysis of such data finds that these disruptive sequences are primarily linked to fast radiative collapses caused by molybdenum impurities. These disruptions have fast timescales, up to a few tens of milliseconds, and we argue that any data-driven disruption prediction algorithm for C-Mod would struggle to predict such cases (at least with the current set of input features) and thus be affected by a high degree of false negatives (missed predictions). The three isolated clusters, identified by red dashed circles in Figure 1, are representative of specific disruption dynamics such as VDEs, impurity accumulations and MHD-driven disruptions. These precursors can be identified through inspection of the specific time series.

## 4. The Hybrid Deep Learning (HDL) disruption prediction framework

Based on our findings about the importance of temporal information, we introduced a Hybrid Deep Learning (HDL) network for time series processing. Figure 2(a) shows the architecture of this network that was used for the cross-machine disruption prediction application reported in this paper. The HDL network consists of two Gated Recurrent Unit (GRU) layers [28], one fully connected layer and three novel Multi-Scale Temporal Convolution (MSTConv) layers, plus the input and the classification layer. The MSTConv layer is inspired from work in machine translation [29], and the detailed structure of one MSTConv layer is shown in Figure 1(b). It consists of six 1-D causal convolution layers [30] with different window lengths L from one to six. The first 1-D convolution layer can only access the current time step $t_0$. The L-th 1-D convolution layer can look at L time steps from $t_{0-L+1}$ to $t_0$. This structure enables different 1-D convolution layers to capture local temporal information at different levels (e.g., 1$^{st}$ order time derivative, 2$^{nd}$ order time derivative, …). The resulting outputs from these six layers are concatenated and then processed through a batch normalization layer [31] and a ReLU (Rectified Linear Unit) activation to develop new features. It is important to highlight that different parts of the HDL architecture serve different purposes: The first two MSTConv layers are used to extract local temporal patterns from the input plasma sequences to form a richer representation of the input space. The following two GRU layers – with their long-term memory capability - can capture the long-range dependencies across different signals in the sequences. Then the subsequent MSTConv and fully connected layers can compress and summarize the output representation from the GRU layers to a 12-dimension latent encoding (dimension of the latent encoding is a tunable parameter) which can be mapped to the output by the classification layer.

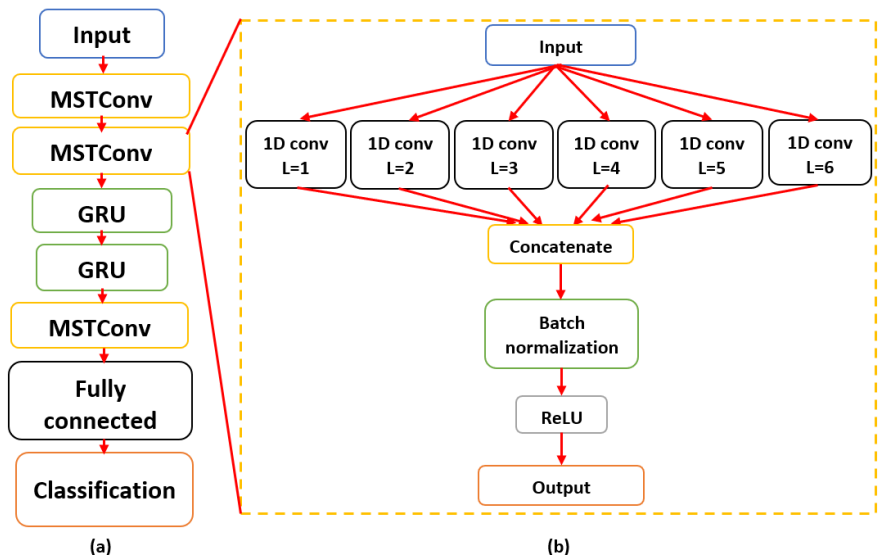

**Figure 2: The HDL architecture (a) and the detailed structure of MSTConv layer (b). Notice that it consists of 6 1-D causal convolution layers with window lengths L from 1 to 6.**

A shot-by-shot testing scheme was designed following [8] to simulate alarms triggered in the Plasma Control System (PCS) using the test shots from different devices. Given an input plasma sequence S which is a 2D matrix consisting of 10 time steps and 12 input features, i.e. a 10×12 matrix, the predictor maps S to a 'disruptivity' between 0 and 1 at the last time step of the sequence; here, 1 is the disruptive class and 0 is the non-disruptive class. During testing, the whole flattop phase of each test shot is subdivided in batches of 10 step sequences, given the HDL architecture design. Each neighbouring testing sequence will have 9 steps overlap, and there are N-9 sequences for a test shot with N steps. If the disruptivity exceeds a pre-set threshold – e.g., 0.5 - at any test time step, the test shot is classified as disruptive and the warning time is recorded for truly disruptive shots, defined as the difference between the alarm time and the final current quench ($t_{dis}$). A successfully detected disruption on C-Mod is shown in Figure 3: under a binary classification scheme, this is regarded as true positive (TP), while false positives (FP) correspond to a false warning, or a healthy plasma being declared to be disruptive. This latter situation can still lead to some machine damage, but on the other hand, being unable to predict a disruption early enough (false negative, FN) is even more costly because it prevents any damage control of disruption consequences. A trade-off can be achieved by adjusting the alarm threshold of the disruptivity, as visually demonstrated by a receiver–operator characteristic (ROC) curve [32]. The area under the ROC curve (AUC) is used as performance metric for the HDL predictor. Throughout the paper, we evaluate the predictive performances on all tokamaks at **50** ms before the disruption event: this is chosen as the minimum warning time to successfully trigger disruption mitigation systems on future devices [33].

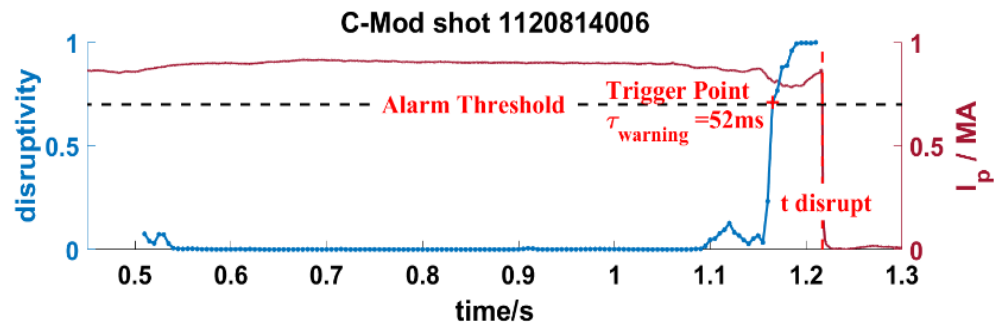

**Figure 3: A successfully detected disruption on C-Mod.**

*4.1 Training technicalities for the HDL model*

Training complex deep neural networks (NN) effectively is a challenging task that involves several technicalities, as also detailed in [9]. Among other things, it is important to address the proper input feature normalization, or to understand which are the tunable parameters that would increase the transferrability of the cross machine predictor, while stabilizing its performance. In the following subsections, we will describe the methods implemented to tackle these challenges for optimally training our deep NN predictor.

*4.1.1 Normalization.* NNs usually need all input features to have similar numerical ranges for all training examples [9, 34]. This makes the use of raw plasma signals as inputs to any NN numerically difficult as different signals have values that can range over many different orders of magnitude. Hence, all 12 signals should be normalized before being used in the network. The normalization should ideally be a common transformation such that it maps a set of signals with the same physical value from different devices to similar numerical values. Different tokamak devices have different operational spaces, spatiotemporal scales, and diagnostics. Moreover, different machines have different event chains toward disruptions, and the most important disruption-relevant physics parameters are different on each machine. Therefore, such a physics-based common transformation is difficult to find, and its extrapolation to ITER is uncertain. However, we find that the best-performing method is to standardize each signal on one machine by its mean and standard deviation across the entire dataset. For each signal on one machine, its normalized form is obtained as follows: $x_{norm} = (x - mean(x))/std(x)$. The normalization parameters of all considered signals on each machine, as well as the common normalization parameters, can be found in the appendix.

The normalization process is independently done on all three machines which implies it is *not* machine-independent: this simple normalization scheme is instead chosen to solve the numerical challenge and leave the generalized signal transformation to the NN. A machine-independent normalization method has also been tested for the three datasets: this normalization standardizes all datasets with a common set of parameters. A performance comparison of HDL predictors using the two normalizations (machine-specific and machine-independent) is shown in figure 4. For the machine specific cases (blue curves), the HDL predictor is trained and evaluated using the training and test sets of each machine normalized by corresponding normalization parameters. For the machine independent cases (red curves), the HDL predictor is trained and evaluated using training and test sets of each machine but normalized by the '*common*' normalization parameters (fixed for all 3 machines), and they give only slightly worse results, hinting at the HDL performance being only weakly dependent on the normalization parameters as long as all signals have proper numerical ranges (approximately -1 to 1).

*4.1.2 **C**ross **M**achine **L**abel **S**moothing (CMLS).* To train a multi-machine predictor, we combine training data from different machines to form a new training set. However, direct mixing of data from various devices can result in a problem: the initial assigned target labels for other devices might not be suitable for the new test device. For example, a certain disruptive sequence from EAST might not be *that* disruptive to C-Mod when compared to C-Mod's disruptive data. Also, a non-disruptive sequence from C-Mod might be slightly unstable to EAST when compared to EAST's non-disruptive data. In other words, we need to take into account the uncertainty associated to running a C-Mod discharge on EAST or on DIII-D and vice versa. To deal with this problem, we choose two smoothing parameters $\varepsilon_1$, $\varepsilon_2$ for each device ($\varepsilon_1$ for non-disruptive examples and $\varepsilon_2$ for disruptive examples) and use these two parameters to modify the target value of the training examples from other machines. When we train the HDL predictor with part of the data from other machines, instead of using their initial (0, 1) target values for non-disruptive examples and disruptive examples, we modify their target values as ($\varepsilon_1$, 1-$\varepsilon_2$). The new target values for non-disruptive examples are $\varepsilon_1$, and the target values for disruptive examples are 1-$\varepsilon_2$. Notice that this modification is only applied to those training examples from other devices; those examples from the test device itself are not modified. We refer to this change in ground-truth target value as the **cross machine label smoothing** technique which we find further improves the cross-machine ability of the HDL predictor (see table 4).

*4.1.3 Hyperparameter tuning and neural networks ensemble.* The HDL disruption predictor has fourteen architectural and two labeling hyperparameters for each device. Guided by our previous numerical experiments on the C-Mod dataset, we roughly scanned the hyperparameter space using a random search for all three machines' data until finding a plateau where any hyperparameter set in this region gives high performance for all three devices. Within this region, changes in hyperparameter choice will only result in minor changes to the model's performance for all three devices. Outside this region, performance on at least one device drops drastically. The hyperparameters of the HDL predictor are therefore selected from the middle of this plateau, and all following qualitative cross-machine conclusions consistently hold for all hyperparameter sets existing in this region. Additionally, our approach includes the adoption of an ensemble of twelve NNs, each one identical in their HDL architecture and tunable hyperparameters but with different initialization seeds. The final prediction comes therefore from an ensemble average. This method is popularly known in the ML community and has been shown to significantly improve the accuracy and stability of the predictor [35-37]. A comprehensive list of tunable hyperparameters for our HDL model can be found in the appendix.

*4.2 HDL performances on the three devices and benchmark with Random Forest*

The HDL predictor successfully achieves state-of-the-art performance on all three test sets when compared to other fully-optimized deep NN disruption predictors [9]. To see this, we trained three HDL predictors (with fixed hyperparameters given in section 4.1.3) and three Random Forest (RF) predictors [6, 8, 19] using the training set of each machine and evaluated their performances on the test set corresponding to that machine. These results are shown in figure 5. To carry on a fair comparison with previous approaches, the RF predictors for each machine are specifically optimized using the corresponding validation set: we carried out a K-fold cross validation procedure together with a parallelized grid search to find the optimal set of time threshold and forest hyperparameters for each machine using a binary classification performance metric called the $F_\gamma$-score [6, 8]. The HDL predictor exceeds RF performances on all three datasets: it triggers fewer false alarms on good discharges and misses fewer real disruptions at the same time, which shows the strong applicability and generalization power of the model. This general improvement on multiple machines seems mainly to come from the advantage of the sequence-based model that is designed for time series processing, as suggested in section 3. Besides its impressive performance, the inference time of our model is short, allowing it to make a prediction in roughly 1ms using an 8-core CPU. This fast and novel model is not only an important step toward the prediction requirement of future devices, but also suggests a powerful conclusion. Given that a common set of model hyperparameters used for three predictors can achieve high performance on all three machines, it suggests that although different devices may have disjoint operational regimes, there seems to exist a common type of discriminant function – same model hyperparameters - capable of separating the disruptive from non-disruptive phases on all these machines.

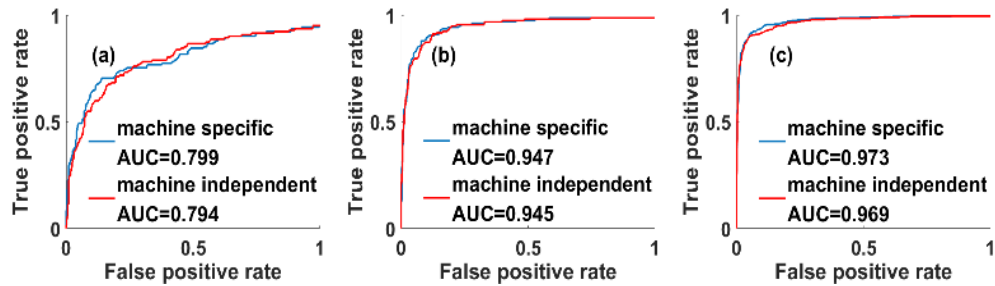

**Figure 4:** The ROC curves from test sets for machine specific normalization (blue) and machine independent normalization (red), for C-Mod (a), DIII-D (b), and EAST (c).

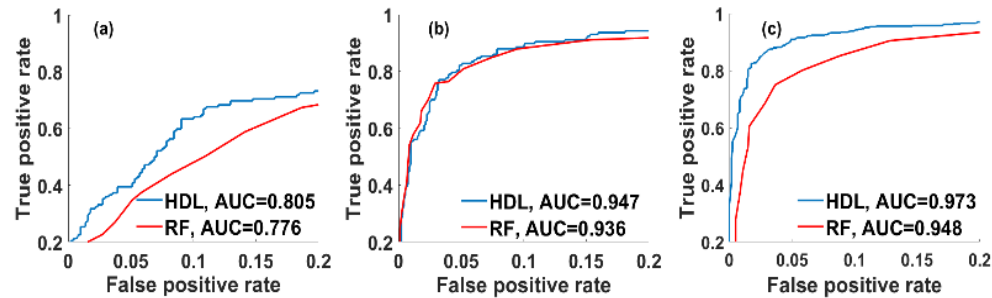

**Figure 5:** The ROC curves from test sets for HDL model and the Random Forest (RF) model, for C-Mod (a), DIII-D (b) and EAST (c). We only show the upper left region of the curves where the predictors have highest performance

## 5. HDL cross machine study on Alcator C-Mod, DIII-D, and EAST data

The availability of a huge amount of experimental data across several tokamaks allows us to design numerical experiments to test transfer learning capabilities of the HDL architecture. Future reactors like ITER cannot tolerate more than a few unmitigated disruptions [1], so we must be able to predict their disruptions given very limited disruptive data from themselves. Expanding from their cross-machine DL-based disruption prediction study, we have designed complete numerical experiments to test transfer learning capabilities of the HDL architecture. Given the availability of a large database of aggregated data from very different tokamaks, it is important to verify *if and how useful* the data from existing devices is to be able to predict unstable plasmas on a new device. In this section, we consider two machines as '*existing machines*' and investigate the effect of their data for the HDL disruption predictor when used on the third machine, chosen as a '*new device*'. We primarily focus on the EAST case (EAST is chosen as '*new device*') in the following section. However, **all following qualitative conclusions are machine-independent**: they always hold no matter which device is selected as the 'new device'. Results regarding the other two cases can be found in the appendix.

### 5.1 Cross-machine prediction performance using the HDL architecture

As first step, we would like to test the cross-machine performance of the HDL model. To do this, we train the HDL network using data from two '*existing devices*' and test its performance on the third unseen '*new device*'. Following the predictors "with a glimpse" or "from scratch" approaches [9, 10], we then add 10 disruptive and 10 non-disruptive discharges from the '*new device*' to the training sets and do indeed observe a boost in the test set performances when using limited data from the target device. In the context of previous cross-machine studies [9,10], our HDL framework shows promising transferrability on these three different devices, and these test results can be found in table 4 (all values reported here are AUCs averaged over the network ensemble).

Beyond performances, we are interested in investigating how data from different existing devices influence predictions of disruptions on a new one, and in particular if any effect can be linked to general, device-independent knowledge. To this aim, we design two further sets of cross-machine numerical experiments. The training set composition for each experiment can be found in table 4.

**Table 3 Cross machine prediction results of HDL**

| Training set | C-Mod+ DIII-D | C-Mod+ DIII-D+ few EAST data | EAST+ C-Mod | EAST+ C-Mod +few DIII-D data | EAST+ DIII-D | EAST+ DIII-D +few C-Mod data |
|---|---|---|---|---|---|---|
| Test set | EAST | EAST | DIII-D | DIII-D | C-Mod | C-Mod |
| HDL Ensemble | 0.788 | 0.819 | 0.622 | 0.741 | 0.564 | 0.605 |
| HDL Ensemble + CMLS | 0.806 | 0.837 | 0.659 | 0.765 | 0.588 | 0.631 |

**Table 4 training set composition of all cross machine experiments using EAST as the '*new machine*'**

| Case NO. | Existing machines (C-Mod+DIII-D) | | New machine (EAST) | |
|---|---|---|---|---|
| | Non-disruptive | Disruptive | Non-disruptive | Disruptive |
| 1 | None | All (692+732) | All (5995) | 20 |
| 2 | None | All | All | None |
| 3 | None | All | 50% (2998) | 20 |
| 4 | None | None | All | 20 |
| 5 | All (2651+4554) | All | All | 20 |
| 6 | All | All | None | None |
| 7 | None | All | All | All (2301) |
| 8 | All | All | All | All |
| 9 | None | None | All | All |
| 10 | None | All | ~33% (1998) | All |
| 11 | ~20% (692+732) | None | ~33% | All |
| 12 | None | None | ~33% | All |

Values in parentheses give the exact numbers of shots.

## 5.2 Cross-machine experiments using limited disruptive data from the 'new device'

The first set of cross-machine experiments was conducted using limited disruptive training shots from the *new device*. The results of these numerical experiments are shown in figure 6(a)-(b). In the first experiment, the disruption predictor is trained on 20 randomly selected disruptive training shots and all non-disruptive training shots from the target *new device,* plus disruptive shots from two other devices (*existing machines*). This combination achieves the best performances on the *new device* test dataset (AUC=0.959, for the EAST case). In the second and third experiment, we first remove all *new device* disruptive shots and then 50% of *new device* non-disruptive shots from the first training dataset, separately. In the fourth experiment, the predictor was trained only using selected *new device* training data (20 disruptive training shots, all non-disruptive training shots), this being our limited data baseline model. In the fifth experiment, we add non-disruptive shots from two other machines to the first training dataset. In the sixth experiment, the predictor is trained only on data from other machines (no *new device* data) and its low performance highlights the importance of non-disruptive data from the target machine. From these numerical experiments, it is possible to draw the following conclusions:

- HDL achieves relatively good performances on a new device if using a few disruptive shots and many non-disruptive shots from the new device plus many disruptive data from existing devices. All components mentioned above are necessary because removing any of them will decrease the performance (cases 1 to 4 in figure 6(a)).

- Non-disruptive data from existing devices is harmful to HDL performance but disruptive data from existing devices improves the predictive power (cases 1, 4, 5 in figure 6(b)).

- Non-disruptive data from the target device can substantially improve the predictive power (case 6 in figure 6(b)).

*5.3 Cross-machine experiments using all disruptive data from the 'new device'*

To further investigate the effect of the class imbalance in the training set, we conducted another set of experiments using all disruptive training shots of the *new device*. The results are reported in figure 7: Again, in the seventh experiment, the disruption predictor is trained on all disruptive and non-disruptive training shots from the *new device*, including disruptive shots from two other machines, and it achieves the best performance on the *new device* test dataset (AUC=0.983, for the EAST case). In the eighth experiment, we add non-disruptive shots from two other machines to the first training dataset. In the ninth experiment, the predictor is trained only on all *new device* training data which is the all data baseline case for comparison. In experiments 10-12 (figure 7(b)), we randomly remove most of *new device* non-disruptive training shots, thus reducing the *new device* non-disruptive training data to be less than *new device* disruptive training data, i.e. an inversely imbalanced situation. The test results from figure 7(a)-(b) point to the following further conclusions:

- Adding disruptive data from existing machines can still slightly improve test performances on the new device even though you have abundant new machine data (cases 7, 9 in figure 7(a)). However, adding non-disruptive data from existing machines is still harmful in this situation (cases 7, 8 in figure 7(a)).

- The effects of disruptive data (positive) and non-disruptive data (negative) do not result from the class imbalance of the new machine dataset, because disruptive data from existing devices continually have positive effects, while the non-disruptive data still have negative effects in the inversely imbalanced situation (figure 7(b)). This difference between disruptive and non-disruptive data is machine independent, i.e. a universal conclusion.

- Also, removing non-disruptive data from the target device will always decrease the test performance no matter how imbalanced the target dataset is (cases 1, 3 in figure 6(a), case 9, 12 in figure 7(b)).

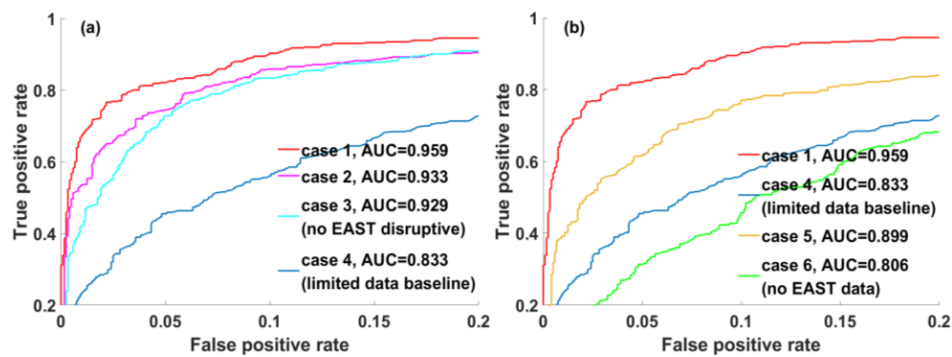

**Figure 6: ROC curves from the EAST test using limited EAST disruptive training data.**

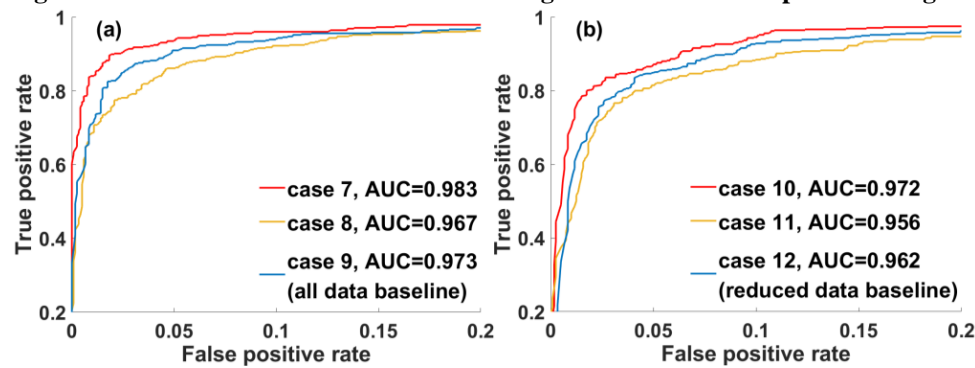

**Figure 7: ROC curves from the EAST test set using all EAST disruptive training data.**

*5.4 Summary of Cross-machine numerical conclusions*

Considering all the conclusions in section 5.2 and 5.3, it is possible to state that knowledge of disruptive data from existing devices improves the performance on the new device while the non-disruptive data seem to have negative effects, which do not result from the label imbalance of training datasets. This suggests that the non-disruptive data are specific to one device, but disruptive data contain some general knowledge about disruptions dynamics that could be transferred to a new device, when using predictive, data-driven models. Indeed, different machines usually have different operational spaces, spatiotemporal scales for physics events and plasma diagnostics [6, 8, 9]. In other words, the distributions of plasma signal can vary significantly from one machine to others. From the data-driven perspective, this further implies that finding a numerical transformation that maps a set of signals from one device to any other devices can be very challenging without incorporating machine-specific information, and this might indeed pose a great challenge when comparing ITER's operational space to all existing devices. Due to these considerations, we are inclined to conclude that non-disruptive data of existing devices are machine-specific and will only decrease the accuracy of the predictive models on the new device when it is directly mixed with data from the target device. Nevertheless, different devices show similar behavior when operating close to a disruption. For example, `li` [6], `v-loop` and `lock mode` [6, 8] signals have been observed to consistently increase on multiple machines when the disruptions are imminent. These universal trends can be well captured by our time sequence based model as general knowledge about disruptions hidden beneath the disruptive data and then help disruption prediction on new devices.

## 6. Summary and future plans

In this paper, we have discussed findings from an explorative data analysis study on C-Mod disruption database using a dimensionality reduction technique called t-SNE and demonstrated that time sequence data can better separate the disruptive and non-disruptive behavior compared to the instantaneous plasma state data (i.e. individual time slices). Based on this conclusion, we have designed a new, powerful disruption prediction algorithm based on Deep Learning and also demonstrated a general, effective way to transfer knowledge from existing devices to new devices which offers guidelines for disruption prediction on new devices using limited disruptive data from target *new* devices. The cross-machine study on Alcator C-Mod, DIII-D, and EAST shows that, given the designed, highly elaborated deep learning architecture, it is not enough to use only data from existing devices to predict disruptions on a new tokamak device. The numerical experiments discussed in Section 5 demonstrate that in addition to data from existing devices, the model's performance greatly improves if both non-disruptive and some disruptive data are included from the target device. In particular, the HDL predictors can reach AUC > 0.95 on EAST if trained including only a small set (20) of disruptive discharges from the target device (EAST), *but* using all available non disruptive information from the target machine.

Furthermore, disruptive and non-disruptive data are found to have a different impact on the cross-machine disruption prediction framework presented in this paper, with the implication that non-disruptive data are machine-specific and the disruptive data contain general knowledge about disruptions. These results are an important milestone for disruption prediction research for next-generation burning plasma reactors, such as ITER. Future efforts will focus on two main topics. Firstly, the precision of our hyperparameter scan is limited by our computing power: Given enough computational resources, we can conduct a fine hyperparameter tuning which might further increase the performance of the predictor and find new insights in the cross-machine study. Secondly, in the future version of the HDL model, we will explore how to directly incorporate device features such as minor radius, major radius, toroidal magnetic field, wall material, etc. The machine-specific characteristics of non-disruptive data suggest that it could be beneficial to mix the device-specific representation with the plasma signal representation to increase the model's expressive power. This may enable us to extract information from machine-specific non-disruptive data and improve the prediction on the new devices. In conclusion, the continuation of this project may indeed contribute to the development of a reliable trigger for ITER's disruption mitigation system, by further refining the guidelines for a robust disruption prediction scheme for yet-to-be-built tokamaks.

**Acknowledgements**

This material is based upon work supported by and the U.S. Department of Energy, Office of Science, Office of Fusion Energy Sciences, using the DIII-D National Fusion Facility, a DOE Office of Science user facility, under Awards DE-FC02-04ER54698 and DE-SC0014264. Additionally, this work is supported by the National MCF Energy R&D Program of China, Grant No. 2018YFE0302100. The HDL architecture reported in the paper was developed using TensorFlow library [38]. Part of the data analysis reported in this paper was performed using the OMFIT integrated modelling framework [39]. The authors are grateful to N. Logan, E. Marmar and R. Buttery for their support and valuable discussions.

**Disclaimer**

**Appendix:**

**1. Details about the sampling rate and time threshold of three disruption warning databases**

Our DIII-D and EAST databases are initially created to have non uniform sampling rates [1]. For DIII-D, all shots (with an average flattop duration of about 3s) are sampled every 25 ms, and additional sampling is done every 2 ms for the 100 ms period before each disruption. For EAST, all shots (average flattop duration ~6s) are sampled every 100 ms (some discharges are 100 s long), and additional sampling is done every 10 ms for the 250 ms period before each disruption. The HDL requires that all input signals are aligned onto a uniform sampling rate: this avoids distortions in the correlations learned among the data due to the increased sampling frequency around the disruption time. Guided by our previous experience about the time scale of disruption dynamics on DIII-D and EAST, we decided to interpolate the DIII-D and EAST databases to have a uniform time base with intermediate sampling rates (10ms for DIII-D, 25ms for EAST), chosen as a trade-off between the additional high-sampling rate before disruption and the low sampling rate for the remaining flattop. For C-Mod, our clustering study combined with further preliminary analysis show that it has considerably faster disruption dynamics than the two other machines. Therefore, we repopulated the database to have the highest possible sampling rate: the EFIT equilibrium reconstruction code determines eventually the maximum achievable sampling rate of 5ms.

Regarding the threshold needed to define the different class labels, disruptive versus stable/non-disruptive sequences, we took into account several different factors, as mentioned in our dataset description section. First, we consulted tokamak operators to obtain the typical range of the unstable phase for disruptive shots on each machine. Then, we studied the distribution of several plasma signals (li, kappa, lock-mode, ip-error-fraction...) to further narrow down such empirical ranges to those thresholds in time for which most distributions start to deviate from their stable counterparts (distributions of non-disruptive phase), as disruptions are approached on these three tokamaks. Finally, for each machine, we scanned all the remaining time threshold "candidate" and chose the one that maximizes the performance on the validation set.

**2. t-Distributed Stochastic Neighbour Embedding (t-SNE) algorithm**

As detailed in [2], t-SNE is a nonlinear dimensionality reduction methods that aim to convert the high-dimensional data set $X = \{x_1, x_2, ..., x_n\}$ into a low-dimensional manifold (usually 2-D or 3-D) set $Y = \{y_1, y_2, ..., y_n\}$ and preserve as much of the significant sturcture of the high dimensional data as possible in its low dimensional representation. The pairwise 'distance' in the low-dimensional map represents the matrix of pairwise similarities between objects: this can be visualized to capture the local stucture of the high-dimensional data but also to reveal global structures, such as the presence of clusters. In the resulting low-dimensional map, nearby points are grouped through similarity criteria. Therefore, two similar objects will appear as two close points while two dissimilar objects will apear as distant point.

The t-SNE algorithm comprises three main stages. First, t-SNE constructs a conditional probability distribution over pairs of high-dimensional objects by converting the high-dimensional Euclidean distances between datapoints into conditional probabilities $p_{j|i}$ that represent similarities and defining the joint probabilities $p_{ij}$ as:

$$p_{j|i} = \frac{\exp\left(-\|x_i - x_j\|^2 / 2\sigma_i^2\right)}{\sum_{k \neq i} \exp(-\|x_i - x_k\|^2 / 2\sigma_i^2)} \quad (1)$$

$$p_{ij} = \frac{p_{j|i} + p_{i|j}}{2n} \quad (2)$$

The variance $\sigma_i$ of the Gaussian that is centered over each high-dimensional datapoints, $x_i$. To find $\sigma_i$ for each Gaussian, t-SNE introduces a hyperparameter 'perplexity' and performs a binary search for the value of $\sigma_i$ that produces a probability distribution $P_i$ with a fixed perplexity. The perplexity is defined as

$$Perp(P_i) = 2^{H(P_i)} \quad (3)$$

Where $H(P_i)$ is the Shannon entropy of $P_i$ measured in bits

$$H(P_i) = -\sum_j p_{j|i} \log_2 p_{j|i} \quad (4)$$

Second, t-SNE calculates a probability distribution over the points in the low dimensional map using Student t-distribution [3] with a single degree of freedom. The joint probabilities $q_{ij}$ are defined as

$$q_{ij} = \frac{\left(1 + \|y_i - y_j\|^2\right)^{-1}}{\sum_{k \neq l}(1 + \|y_k - y_l\|^2)^{-1}} \quad (5)$$

Then, t-SNE will minimize the Kullback-Leibler divergence (KL divergence) between the two distributions with respect to the low dimensional representation using gradient descent. The gradient of the KL divergence between high-dimensional distribution $P$ and low dimensional distribution $Q$ is given by

$$\frac{\partial C}{\partial y_i} = 4 \sum_j (p_{ij} - q_{ij})(y_i - y_j)\left(1 + \|y_i - y_j\|^2\right)^{-1} \quad (6)$$

| **Algorithm**: t-Distributed Stochastic Neighbor Embedding |
|---|
| **Data**: data set $X = \{x_1, x_2, ..., x_n\}$ |
| cost function parameter: perplexity *Perp,* |
| optimization parameters: number of iterations *T*, learning rate η, momentum α(t) |
| **Result**: low-dimensional data representation $Y = \{y_1, y_2, ..., y_n\}$ |
| **begin** |
|     compute pairwise similarites $p_{ij}$ with *Perp* (using Eq. 1&2) |
|     sample initial solution $Y^0 = \{y_1, y_2, ..., y_n\}$ from $N(0, 10^{-4}I)$ |
|     **for** t=1 **to** *T* **do** |
|         compute low dimensional similarities $q_{ij}$ (using Eq. 5) |
|         compute gradient g (using Eq. 6) |
|         set $Y^t = Y^{t-1} + \eta g + \alpha(t)(Y^{t-1} - Y^{t-2})$ |
|     **end** |
| **end** |

## 3. t-SNE clustering results for EAST and DIII-D databases

*3.1 EAST clustering result*

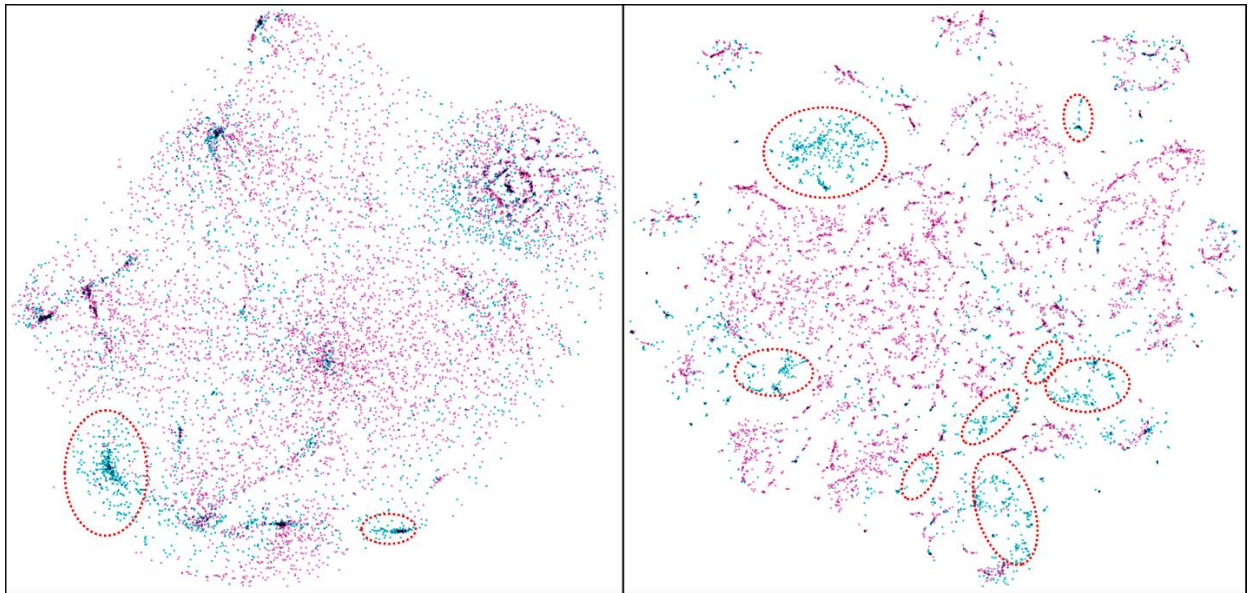

**Figure 1: t-SNE clustering for visualizing EAST data. On the left, t-SNE is performed on individual disruptive (cyan) and non-disruptive (magenta) time slices while on the right t-SNE is performed on 10-step disruptive (cyan) and non-disruptive (magenta) sequences. The colouring is done a-posteriori.**

    The figure above shows the t-SNE algorithm applied to time slice data (left) and aggregated sequence data (right) for EAST disruption warning database. In the left subplot, each magenta point represents a randomly sampled time slice (a 12-element array composed by the 12 plasma signals from table 1 in the main text) from the flattop of a non-disruptive shot, while each cyan point represents a time slice randomly sampled from the last 300 ms of a disruptive shot. On the right, each cyan point represents a 10-step (a 10 x 12 element matrix) sequence randomly sampled from the last 300 ms of a disruptive shot while each magenta point represents a 10-step sequence randomly sampled from the flattop of a non-disruptive shot. The colouring of each datapoint in the plots is done *a-posteriori*, therefore not provided during the training process characterizing the t-SNE as an unsupervised clustering technique.

    As can be seen from two subplots, the clustering of individual time slices separates one major disruptive cluster and one minor disruptive cluster from the bulk of non-disruptive samples. But there are more than half of cyan points still mixed with the magenta points. On the contrary, the clustering of time-sequence data successfully separates one major disruptive cluster as well as several smaller disruptive clusters from the magenta points. There are only a few disruptive samples (<15%) remaining mixed with the non-disruptive samples. Again, migrating from time slice clustering to time sequence clustering gives considerably better separation of disruptive and non-disruptive regions.

*3.2 DIII-D clustering result*

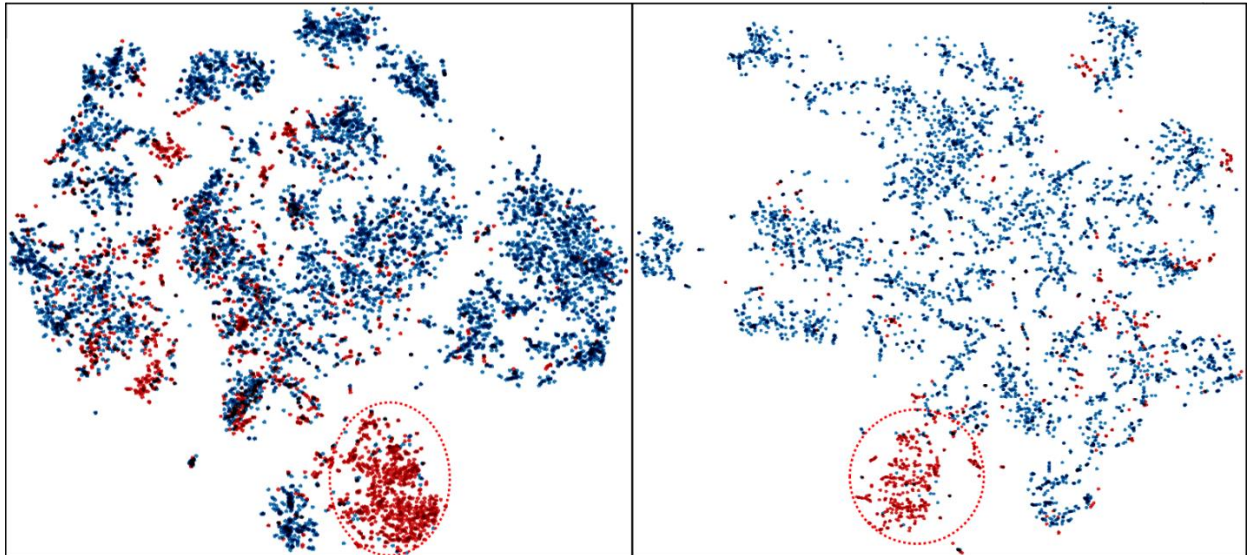

**Figure 2: t-SNE clustering for visualizing DIII-D data. On the left, t-SNE is performed on individual disruptive (red) and non-disruptive (blue) time slices while on the right t-SNE is performed on 10-step disruptive (red) and non-disruptive (blue) sequences. The colouring is done a-posteriori.**

The figure above shows the t-SNE algorithm applied to time slice data (left) and aggregated sequence data (right) for DIII-D disruption warning database. In the left subplot, each blue point represents a randomly sampled time slice (a 12-element array composed by the 12 plasma signals from table 1 in the main text) from the flattop of a non-disruptive shot, while each red point represents a time slice randomly sampled from the last 200 ms of a disruptive shot. On the right, each red point represents a 10-step (a 10 x 12 element matrix) sequence randomly sampled from the last 200 ms of a disruptive shot while each blue point represents a 10-step sequence randomly sampled from the flattop of a non-disruptive shot. The colouring of each datapoint in the plots is done *a-posteriori*, therefore not provided during the training process characterizing the t-SNE as an unsupervised clustering technique.

The DIII-D case is slightly different from C-Mod and EAST cases. The clustering of individual time slices successfully separates one major disruptive cluster from non-disruptive samples which counts for ~70% of all disruptive samples. The clustering of time sequences also separates one major cluster from non-disruptive samples which counts for >80% of all disruptive samples. In the DIII-D case, we find that migrating from time slice clustering to sequence clustering still gets better separation, but only marginally. And the clustering of time slice data already gives good results. Further analysis shows that both major clusters in the two clustering schemes are related to locked mode-induced disruptions. These results suggest that the features used to describe DIII-D data are well-suited to identify locked mode disruptions, independently of whether the data is described in terms of time slices or sequences. This reconnects with previously documented literature [1, 4] where various time slice-based models (e.g. random forest) work well to describe DIII-D instantaneous plasma state and the sequence-based HDL can only slightly improve the performance over the results of the optimized random forest predictor (see figure 5(b)). However, in order to detect various other types of disruptions, we argue that a more suitable approach is the study of time sequences. In relation to C-Mod and EAST, DIII-D data seems to have some special characteristics, with a high prevalence of one particular class of disruptions (locked mode related disruptions), that makes both predictive approaches (time sequence vs time slices) comparably effective.

## 4. Normalization parameters of each signal

**Table 1: Normalization parameters of each signal on three machines and the common normalization set**

| Plasma signal | Mean EAST | Std EAST | Mean DIII-D | Std DIII-D | Mean C-Mod | Std C-Mod | Mean Common | Std Common |
|---|---|---|---|---|---|---|---|---|
| `ip-error-fraction` | -0.002 | 0.021 | -0.016 | 0.055 | -0.002 | 0.043 | -0.007 | 0.040 |
| `locked-mode-proxy` | 0.002 | 0.004 | $1.546 \cdot 10^{-4}$ | $3.503 \cdot 10^{-4}$ | $7.495 \cdot 10^{-4}$ | $4.364 \cdot 10^{-4}$ | $9.680 \cdot 10^{-4}$ | $1.596 \cdot 10^{-3}$ |
| `Greenwald-fraction` | 0.437 | 0.327 | 0.407 | 0.184 | 0.261 | 0.132 | 0.368 | 0.214 |
| `lower-gap` | 0.160 m | 0.032 m | 0.170 m | 0.082 m | 0.056 m | 0.016 m | 0.129 m | 0.043 m |
| `z-error-proxy` | 0.007 m | 0.022 m | $9.114 \cdot 10^{-4}$ m | 0.006 m | $-8.459 \cdot 10^{-7}$ m | 0.002 m | $2.637 \cdot 10^{-3}$ m | 0.010 m |
| `kappa` | 1.630 | 0.114 | 1.768 | 0.105 | 1.618 | 0.092 | 1.672 | 0.104 |
| `betap` | 0.689 | 0.380 | 0.826 | 0.501 | 0.239 | 0.184 | 0.585 | 0.355 |
| `radiated-fraction` | 0.138 | 0.347 | 0.516 | 1.237 | 0.369 | 0.952 | 0.341 | 0.845 |
| `rotating-mode-proxy` | 0.005 | 0.011 | $6.823 \cdot 10^{-5}$ | $1.554 \cdot 10^{-4}$ | $0.678\ S^{-1}$ | $1.062\ S^{-1}$ | 0.228 | 0.358 |
| `v-loop` | 0.430 V | 0.861 V | 0.293 V | 0.936 V | -0.329 V | 1.774 V | 0.131 V | 1.190 V |
| `q95` | 6.009 | 1.275 | 4.860 | 1.417 | 4.422 | 0.943 | 5.097 | 1.212 |
| `li` | 1.187 | 0.228 | 1.015 | 0.215 | 1.404 | 0.172 | 1.202 | 0.205 |

## 5. Recommended hyperparameters for HDL framework

**Table 2: Explanation of all hyperparameters and their recommended values**

| Hyperparameter | Explanation | Best Value |
|---|---|---|
| η | Learning rate | $5 \times 10^{-4}$ |
| beta2 | The exponential decay rate for the second-moment estimates | 0.970 |
| $N_{GRU}$ | Number of **G**ated **R**ecurrent **U**nit (GRU) Layers | 2 |
| $n_{cells-1}$ | Number of GRU cells in layer 1 | 130 |
| $n_{cells-2}$ | Number of GRU cells in layer 2 | 90 |
| $n_{batch}$ | Batch size | 300 |
| Target | Type of target function | Negative Log-Likelihood (NLL) |
| $N_F$ | Number of convolutional filters in the 1-D causal convolution sublayers of each **M**ulti-**S**cale **T**emporal **Conv**olution (MSTConv) layer | 10 |
| Optimizer | Stochastic optimization scheme | Adam |
| Dropout | Dropout probability | 0.1 |
| L2 regularization | Weight regularization of all weight | $1 \times 10^{-3}$ |
| $n_{epoch}$ | Number of training epochs | 32 |
| $d_{latent}$ | Dimension of final latent representation | 12 |
| $N_{MSTConv}$ | Number of MSTConv Layers | 3 |

| $\varepsilon_{1\text{-C-Mod}}$ | Smoothing parameter $\varepsilon_1$ when test on C-Mod | 0.00 |
|---|---|---|
| $\varepsilon_{2\text{-C-Mod}}$ | Smoothing parameter $\varepsilon_2$ when test on C-Mod | 0.08 |
| $\varepsilon_{1\text{-DIII-D}}$ | Smoothing parameter $\varepsilon_1$ when test on DIII-D | 0.00 |
| $\varepsilon_{2\text{-DIII-D}}$ | Smoothing parameter $\varepsilon_2$ when test on DIII-D | 0.05 |
| $\varepsilon_{1\text{-EAST}}$ | Smoothing parameter $\varepsilon_1$ when test on EAST | 0.09 |
| $\varepsilon_{2\text{-EAST}}$ | Smoothing parameter $\varepsilon_2$ when test on EAST | 0.00 |

## 6. Cross-machine numerical experiments using C-Mod and DIII-D as target devices

### 6.1 Cross-machine experiments testing on C-Mod

In these experiments, we consider DIII-D and EAST as 'existing machines' and investigate the effect of their data for the HDL disruption predictor when used on C-Mod, chosen as a 'new device'. The numerical results are shown in supplementary figure 3-4 which consistently support our previous cross-machine conclusions, to be found in Section 5, cross-machine study. Notice that, given enough C-Mod data (cases in figure 4), continually adding data from existing machines (DIII-D, EAST) can only slightly change the performance on C-Mod. This fact implies C-Mod disruptions are quite different from disruptions on other existing devices which agrees with our findings from the clustering study. Therefore, with enough data from C-Mod, data from other existing devices barely help the disruption prediction on C-Mod. The detailed training set composition for each case can be found in table 3.

**Table 3 training set composition of all cross machine experiments using C-Mod as the '*new device*'**

| Case NO. | Existing machines (DIII-D+EAST) | | New machine (C-Mod) | |
|---|---|---|---|---|
| | Non-disruptive | Disruptive | Non-disruptive | Disruptive |
| 1 | None | All (732+2301) | All (2651) | 20 |
| 2 | None | All | All | None |
| 3 | None | All | 50% (1326) | 20 |
| 4 | None | None | All | 20 |
| 5 | All (4554+5995) | All | All | 20 |
| 6 | All | All | None | None |
| 7 | None | All | All | All (692) |
| 8 | All | All | All | All |
| 9 | None | None | All | All |
| 10 | None | All | ~25% (662) | All |
| 11 | ~29% (732+2301) | None | ~25% | All |
| 12 | None | None | ~25% | All |

Values in the parentheses give the exact numbers of shots.

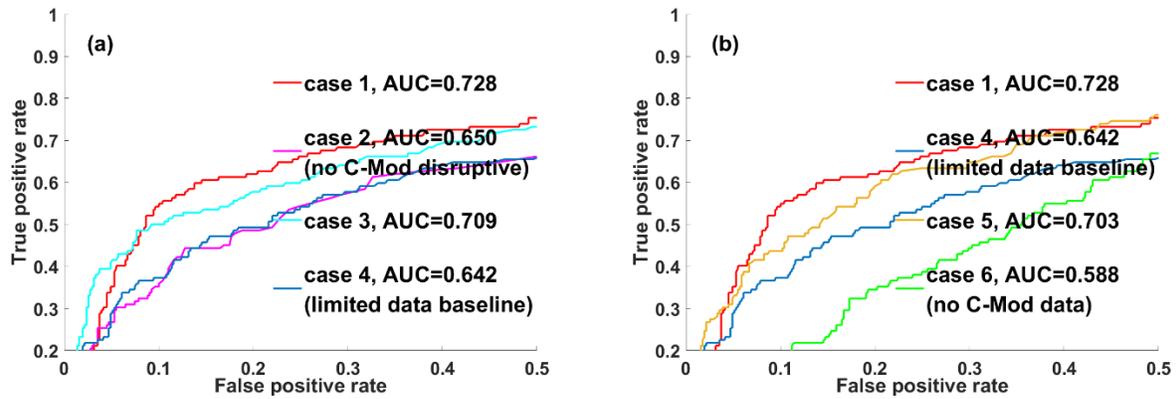

**Figure 3:** ROC curves from the C-Mod test set using limited C-Mod disruptive training data.

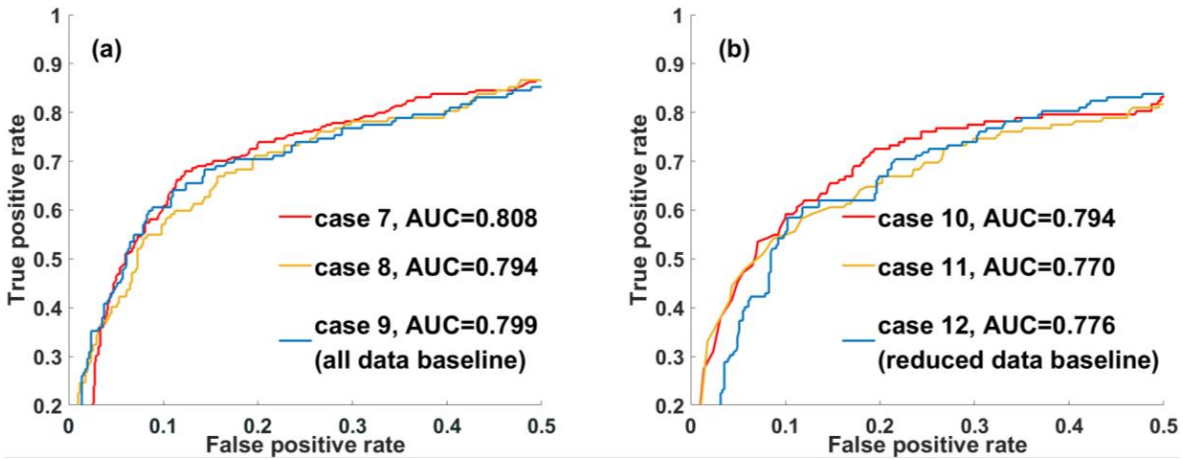

**Figure 4:** ROC curves from the C-Mod test set using all C-Mod disruptive training data.

*6.2 Cross-machine experiments testing on DIII-D*

In these experiments, we consider C-Mod and EAST as 'existing machines' and investigate the effect of their data for the HDL disruption predictor when used on DIII-D, chosen as a 'new device'. The numerical results are shown in supplementary figure 5-6 which consistently support our previous cross-machine conclusions, to be found in Section 5, cross-machine study. And the detailed training set composition for each case can be found in table 4.

**Table 4** training set composition of all cross machine experiments using DIII-D as the '*new device*'

| Case NO. | Existing machines (C-Mod+EAST) | | New machine (DIII-D) | |
|---|---|---|---|---|
| | Non-disruptive | Disruptive | Non-disruptive | Disruptive |
| 1 | None | All (692+2301) | All (4554) | 20 |
| 2 | None | All | All | None |
| 3 | None | All | 50% (2277) | 20 |
| 4 | None | None | All | 20 |
| 5 | All (2651+5995) | All | All | 20 |
| 6 | All | All | None | None |
| | | | | |
| 7 | None | All | All | All (732) |
| 8 | All | All | All | All |
| 9 | None | None | All | All |
| 10 | None | All | ~15% (700) | All |
| 11 | ~35% (692+2301) | None | ~15% | All |
| 12 | None | None | ~15% | All |

Values in the parentheses give the exact numbers of shots.

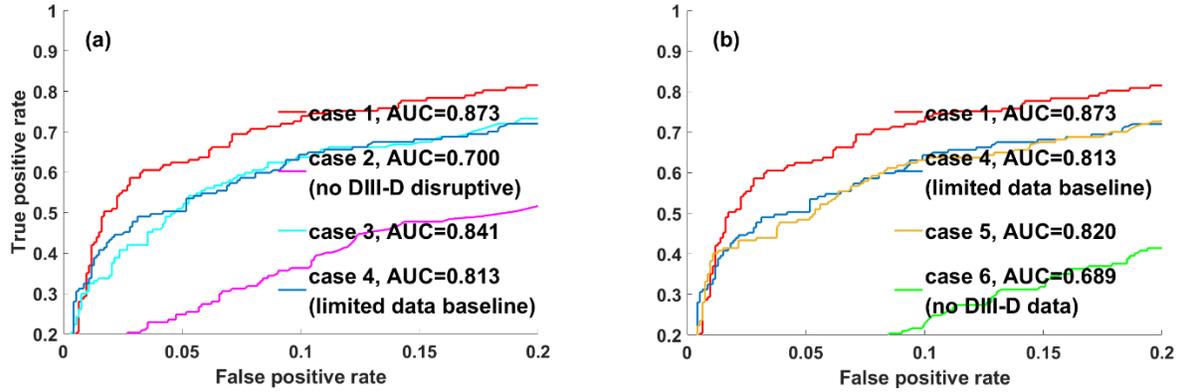

**Figure 5:** ROC curves from the DIII-D test set using limited DIII-D disruptive training data.

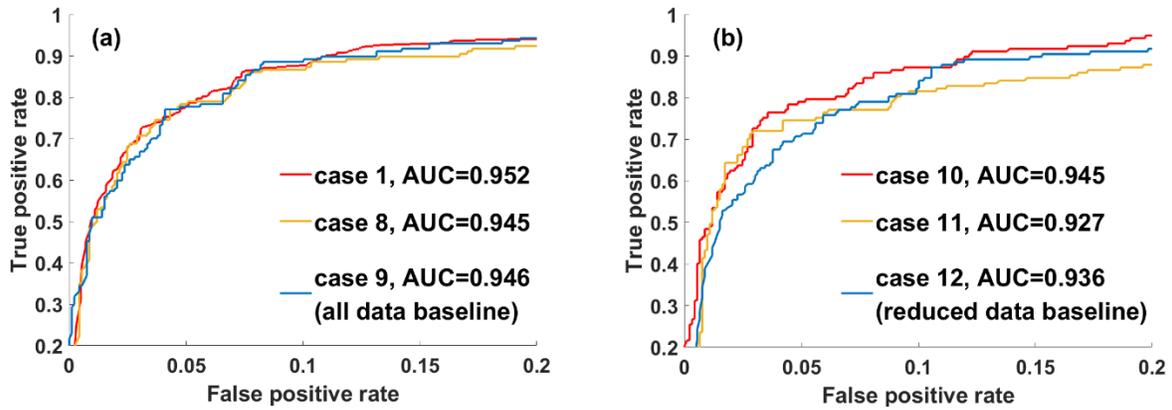

**Figure 6:** ROC curves from the DIII-D test set using all DIII-D disruptive training data.